\documentclass[aps,pra,floatfix,preprint]{revtex4-1}

\usepackage{amsmath, amssymb}
\usepackage{mathrsfs}
\usepackage{graphicx}
\usepackage{hyperref}

\newcommand{\nn}{\nonumber}
\newcommand{\mycomment}[1]{}
\newcommand{\hlab}[1]{\mathscr{H}_{#1}}
\newcommand{\hint}[1]{H_{#1}}

\begin{document}
	
	\title{Optimal enhancement of the Overhauser and Solid Effects within a unified framework}
	\author{Sarfraj Fency}
	\email{smjf21ip029@iiserkol.ac.in}
	\affiliation{Department of Physical Sciences,
		Indian Institute of Science Education and Research Kolkata, Mohanpur 741246, India}
	\author{Rangeet Bhattacharyya}
	\email{rangeet@iiserkol.ac.in}
	\affiliation{Department of Physical Sciences,
		Indian Institute of Science Education and Research Kolkata, Mohanpur 741246, India}
	
\begin{abstract}
	The Overhauser effect (OE) and the Solid effect (SE) are two Dynamic Nuclear Polarization
	techniques. These two-spin techniques are widely used to create nonequilibrium nuclear
	spin states having polarization far beyond its equilibrium value. OE is commonly
	encountered in liquids, and SE is a solid-state technique.  Here, we report a single
	framework based on a recently proposed quantum master equation, to explain both OE and SE.  To this end, we use a
	fluctuation-regularized quantum master equation that predicts dipolar relaxation and
	drive-induced dissipation, in addition to the standard environmental dissipation channels.
	Importantly, this unified approach predicts the existence of optimal microwave drive amplitudes
	that maximize the OE and SE enhancements. We also report optimal enhancement regime for
	electron-nuclear coupling for maximal enhancement.
\end{abstract}
	
\maketitle

\section{Introduction}
Hyperpolarization is used to prepare nuclear spin ensembles in strongly nonequilibrium
states, producing polarization that exceeds thermal limits \cite{eills2023spin}. Dynamic
nuclear polarization (DNP) is one of the most established and conceptually minimal
hyperpolarization mechanisms \cite{lily2017, corzilius2020high}. In DNP, polarization is
transferred from an electron spin to a hyperfine-coupled nuclear spin under microwave
irradiation \cite{overhausser, carver1953polarization, jeffries57, odehnal1959nuclear,
goldman1963dynamic, jeffries1965dynamic, abraham1957gamma, abragam1958new}. Although DNP
involves only a few degrees of freedom, even the simplest electron–nuclear system exhibits
rich frequency-dependent dynamics, making DNP a typical example of nonequilibrium spin
physics in the solid state.

Traditionally, two distinct DNP mechanisms are identified when a single electron and a
single nucleus are considered, the Overhauser effect (OE) \cite{overhausser} and the Solid
effect (SE) \cite{jeffries57, odehnal1959nuclear}.  In the OE, microwave irradiation
saturates the electron spin levels, and electron–nuclear cross-relaxation leads to nuclear
polarization enhancement. On the other hand, in the Solid Effect, microwave irradiation at
forbidden electron–nuclear transitions directly drives polarization transfer, resulting in
nuclear polarization enhancement. As such, by controlling the microwave irradiation
frequency one can selectively use OE or SE. 

Usually, the mechanisms involved in OE and SE are treated as separate
phenomena involving different driving frequencies and derived using different theoretical
assumptions \cite{hwang1967phenomenological, schaefer1973distributions, tropp1980dipolar,
thurber2014mechanisms}. While both originate from the same electron–nuclear interactions
under microwave driving, their conventional descriptions emphasize different theoretical
treatments, obscuring their common physical origin and mutual connection.


Typically, the theories of DNP use rate equations \cite{leifson1961dynamic,
schmugge1965high, wollan1976dynamic, abragam1978principles, farrar2001mechanism,
ardenkjaer2008dynamic, smith2012solid}, Bloch equations \cite{thompson1964dynamic,
sezer2023solid, sezer2023non}, spin temperature theory \cite{goldman1963dynamic,
de1976dynamic}, spin density operator formalism \cite{hovav2010theoretical} or some
phenomenological models \cite{hwang1967phenomenological, wollan1976dynamic}. Despite the
success of these models, there is a lack of single model that describes both the two-spin 
mechanisms of DNP.

To this end, we use a fluctuation-regularized quantum master equation (FRQME) to describe
two-spin mechanisms of DNP \cite{frqme}. This recently-proposed formalism takes a
coarse-grained approach to arrive at a time-local quantum master equation for a quantum
system by taking into account the fluctuations in its local environment. The important
feature of the FRQME is the inclusion of closed-form second-order terms from the
perturbing Hamiltonians.  As such, we have dipolar relaxation terms from the dipolar
couplings and drive-induced dissipation (DID) terms from external drives \cite{chakrabarti2018non}. Having these
terms in the master equation help explain OE (from the dipolar cross-relaxation) and SE
(from the drive and the coupling). We note that FRQME had been successfully used in
quantum control \cite{chanda2023optimal}, quantum dynamics of prethermal regimes \cite{saha2023cascaded, saha2024prethermal},
quantum foundations \cite{chanda2021emergence, das2024irreversibility}, quantum sensing \cite{chatterjee2024improved}, NMR \cite{saha2022time},
etc.


In this letter, we show that FRQME describes the DNP dynamics as we vary the microwave
irradiation frequency from resonant Larmor (OE) to off-resonance forbidden transitions
(SE).  Moreover, owing to the presence of DID, we find the existence of 
optimal microwave powers and
coupling strengths that maximize nuclear polarization enhancement. We demonstrate that
polarization transfer exhibits a non-monotonic dependence on driving strength and coupling
strength, reflecting a competition between external driving and environmental dissipation.
Consequently, we obtain optimal conditions for both the DNP mechanisms.


\section{The model} \label{section: model}

We consider a simple two-spin system comprising of a dipolar-coupled
spin-half nuclear spin and a spin-half electron spin, with a static Zeeman Hamiltonian
$\hlab{\circ} =  \omega_e S_z + \omega_n I_z$, where, $\omega_e$ and $\omega_n$ are Larmor
frequencies of the electron and the nucleus, respectively. $S_z = \mathbb{I} \otimes \frac{1}{2}\sigma_z $ 
and $I_z = \frac{1}{2} \sigma_z \otimes \mathbb{I}$ are electron and nuclear spin
operator components along the $z$ direction. The relevant part of the coupling between the
electron and nucleus is $\hlab{\text{\tiny DD}} = \omega_{\text{\tiny d}}\left( C I_{+} S_{z} 
+ C^{*}I_{-} S_{z} \right),$ where, $C = - \frac{3}{4} \sin 2\theta e^{-i\phi}$, $\theta$
and $\phi$ represent the Euler angles of the dipolar vector
\cite{smith1992hamiltonians}. The microwave drive on the electron is described by
$\hlab{\text{drive}} = \omega_1 \left( S_{x} \cos{\omega_{\mu}t} + S_{y} \sin{
\omega_{\mu}t} \right)$, where, $\omega_1$, and $\omega_{\mu}$ are the drive's amplitude and
carrier frequency, respectively. We have chosen a circularly polarized form for
simplicity. Further, we assume that both the electron and the nucleus
are connected to their respective local environments. The coupling with the local
environment is conveniently described by a Jaynes-Cummings type Hamiltonian
$\hlab{\text{\tiny EL}} = \omega_{\text{\text{\tiny EL}}} \left( S_{+} L^{e}_{-} +
S_{-} L^{e}_{+}\right)$ for the electron, and $\hlab{\text{\tiny NL}} = 
\omega_{\text{\text{\tiny NL}}} \left( I_{+} L^{n}_{-} +
I_{-} L^{n}_{+}\right)$. Here, $L$ operators represent local environment's ladder
operators. The local environment is modelled as 
simple resonant two-level systems for the electron and the nucleus, at the same
temperature. As such, a bath static Hamiltonian is introduced as
$\hlab{\circ}^{\text{env}} = \omega_{e}L_z^e + \omega_n L_z^n$ and a density matrix
$\rho_L^{eq} = exp(-\beta \hlab{\circ}^{\text{env}})/Z$, where $Z$ is the partition
function and $\beta$ is the inverse temperature. We note that for this choice of the
operators, we have $\text{Tr}_{\text{L}}\{L^{e,n}_{\pm}\rho_L^{eq}\} = 0$. Hence, the
leading-order contribution from
the coupling with local environments in the master equation is in the second-order.
Next, we describe the dynamics of the system using FRQME.

We shall assume, as is customary in FRQME, that these local environments experience
thermal fluctuations. We assume that the local environments of the electron and the
nucleus are part of the large bath at temperature $T$.  
We also note that the FRQME requires that $\tau_c$ is much shorter than
the system's typical timescale of evolution. In our system, it is expected that bath
fluctuations are rapid compared to the \emph{slow} evolution of the system and hence this
requirement is adequately met.
For convenience of the calculations, we shall describe the dynamics in
the drive frame of the microwave and the laboratory frame for the nuclear part. We note that FRQME is usually described in the
interaction representation of the system with respect to $\hlab{\circ} + \hlab{\circ}^{\text{env}}$. 
As such, we use additional transformation
$\exp(iHt)$, where $H = (\omega_\mu-\omega_e)S_z - \omega_n I_z$, to arrive at this frame. 
Following the transformations, FRQME in the microwave drive frame assumes the form,
\begin{eqnarray} \label{frqme_drive}
\dot{\rho}_{\text{s}} &=& - i [H_{\text{shift}}   + \hint{\text{eff}}(t), \, \rho_s] - \int_{0}^{\infty} \kern-7pt d\tau \:  e^{-\tau/\tau_{c}}\,  
[\hint{\text{eff}}(t), \, [\hint{\text{eff}}(t- \tau), \, \rho_s]]^{\text{sec}} + \mathscr{D}_{e}(\rho_s) + \mathscr{D}_{n}(\rho_s) \nn \\
\end{eqnarray}

Here, $H_{\text{shift}} = \delta \omega S_z +\omega_n  I_z $ is the shift Hamiltonian with
the offset $\delta \omega = \omega_{e} - \omega_{\mu}$, $H_{\text{eff}} = H_{\text{\tiny
DD}} + H_{\text{\tiny drive}}$ is the sum of the drive and dipolar coupling in the drive
frame of the electron. The upright $H$ is indicative of the drive frame as opposed to
lab-frame's cursive $\hlab{}$. $\rho$ is the density matrix, $\rho_{\text{s}} =
\text{Tr}_{\text{\tiny L}}(\rho)$ is the system's density matrix in the drive frame of the
electron and $\tau_{c}$ is the environmental correlation time. 
In the coarse-grained
approach of deriving FRQME, the rapidly oscillating terms from the double commutator
vanishes and only slow or constant terms survive. The superscript ``sec" indicates that
these \emph{secular} pairs are to be retained in the calculation
\cite{cohen2024atom, frqme}.
It is important to note that
the Hamiltonian at $(t-\tau)$ in the inner commutator is also converted to the new frame
using the same transformation. $\mathscr{D}_{e}(\rho_s)$ and $\mathscr{D}_{n}(\rho_s)$ are
the Lindbladian dissipators for the electron and the nucleus, respectively. The explicit
form of the electron dissipator is given by, $\mathscr{D}_{e}(\rho_s) =
\omega_{EL}^2\tau_c[e^{\beta \hbar\omega_e/2}(S_{-}\rho_s S_{+}
-\frac{1}{2}\{S_{+}S_{-},\rho_s\})/Z + e^{-\beta\hbar\omega_e/2}(S_{+}\rho_s S_{-}
-\frac{1}{2}\{S_{-}S_{+},\rho_s\})/Z]$, where $Z$ is the partition function of the model
electron bath. The nuclear part has a similar form.

We note that the new terms in this formalism is the inclusion of the second-line integral
in the eq. (1). The dipole-dipole auto term in this integral is responsible for the OE
transfers. In SE, the single commutator and the dissipators in the third line play the
major roles. However, the presence of the drive-drive cross term in the second line show
deleterious effects on the dynamics for large values of $\omega_1$.



%

\section{Results} \label{section:results} The equation of motion in eq.
(\ref{frqme_drive}), due to its complexity, has been solved numerically using the
parameters routinely used in experiments \cite{becerra1993dynamic, palani2023dynamic,
kavtanyuk2025achieving}.  A low temperature was assumed ($\sim 65$K), and the equilibrium
population was calculated using the Boltzmann factor corresponding to this temperature.
Following standard practice, we define the polarization enhancement ($\epsilon$) as the
ratio of the nuclear polarization at the steady-state, $P^{ SS}$, to its equilibrium
value, $P^{eq}$, 

\begin{equation}
\epsilon = \frac{P^{SS}}{P^{eq}} = \frac{\text{Tr}_{\text{s}}\{\rho_{\text{s}}(t \rightarrow \infty)
I_z\}}{\text{Tr}_{\text{s}}\{\rho_{\text{s}}(t = 0) I_z\}}
\end{equation}

\begin{figure}[t!]
\centering
\includegraphics[scale=0.4]{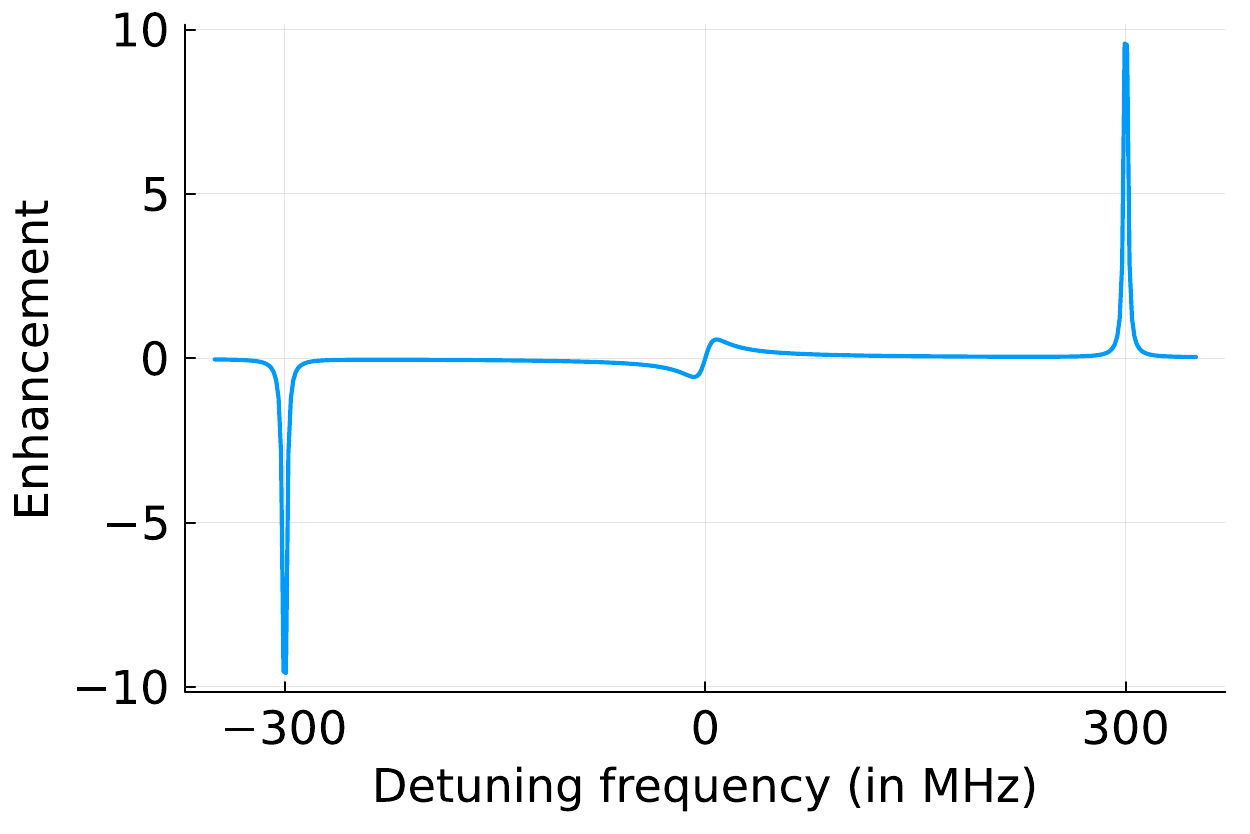}
\caption{The figure shows four peaks as we sweep the drive frequency and plot the
corresponding enhancement in nuclear polarization. The
peaks at $\Delta \omega/2\pi = \pm 300$ MHz collectively describe the solid-effect
mechanism of DNP, while the peaks around $\Delta \omega/2\pi = 0$ MHz describe the
well-known Overhauser effect. The parameters used are $\omega_{n}/2\pi = 300$\,MHz,
$\omega_{e} =  10^{3} \times \omega_{n}$, $\omega_{\text{d}}/2\pi =  3$\,MHz, $\theta =
\pi/3$, $\phi = 0$, $\omega_{1}/2\pi =  8$\,MHz, $\omega_{\text{\tiny EL}}/2\pi= 10$\,MHz,
$\omega_{\text{\tiny NL}}/2\pi = 0.2$\,MHz, and $\tau_{c} = 1$\,ns.}
\label{fig:Spectrum}
\end{figure}

We sweep the microwave irradiation frequency within the frequency range $\omega_e \pm
(\omega_n + 2\pi\times 50)$\,M rad s$^{-1}$ to cover the forbidden transitions. For each
frequency value, we solve eq. (\ref{frqme_drive}) to calculate the steady-state nuclear
polarization and plot the corresponding enhancement on the y-axis of fig.
\ref{fig:Spectrum}, which shows four peaks, in agreement with the experimentally observed
spectrum for two-spin DNP \cite{hu2011quantum}. The peak at detuning $\Delta \omega =
\omega_{\mu} - \omega_e = \omega_n$ is due to double quantum transitions, and the peak at
$\Delta \omega =  -\omega_n$ is due to zero quantum transitions. These two resonant peaks
collectively describe the solid effect mechanism \cite{jeffries57}. We note that the
intermixing of the energy levels of electron and nuclear subsystems by dipolar coupling
allows the so-called forbidden transitions \cite{de1976dynamic, lily2017}. We attribute
the dispersive profile around zero detuning to the well-known Overhauser effect, whose
peaks are shifted due to the intermixing of energy levels \cite{overhausser}.

\begin{figure}[t]
\centering
\includegraphics[scale=0.85]{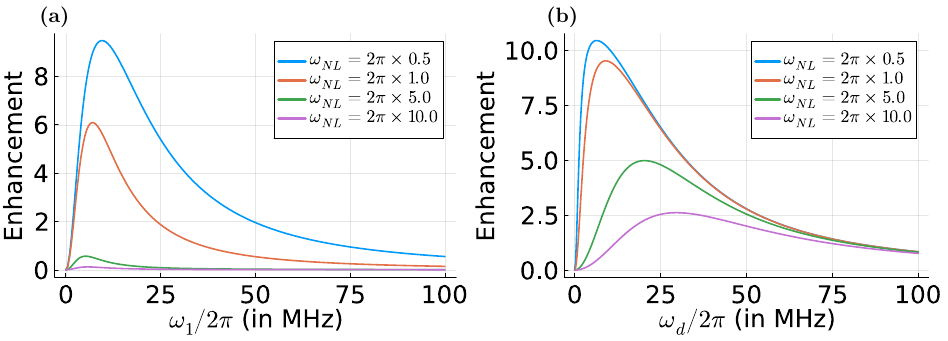}
\caption{The optimal behavior of enhancement in nuclear polarization is plotted as we vary
(a) drive strength, (b) dipolar coupling strength for different nuclear subsystems'
coupling with the local environment. The parameters used are $\omega_{n}/2\pi = 300$\,MHz,
$\omega_{e} =  10^{3} \times \omega_{n}$, $\omega_{\text{d}}/2\pi =  3$\,MHz (for a),
$\theta = \pi/3$, $\phi = 0$, $\omega_{1}/2\pi =  8$\,MHz (for b), $\omega_{\text{\tiny
EL}}/2\pi= 10$\,MHz, and $\tau_{c} = 1$\,ns. Here, $\omega_{\text{\tiny NL}}$ is in units
of \,M\,rad\,s$^{-1}$.}
\label{fig:w1-wd}
\end{figure}

So, our numerical investigation 
reveals that the system shows an optimal behavior with respect to multiple
parameters. We vary $\omega_1$ and plot the corresponding enhancement in nuclear
polarization for different $\omega_{\text{\tiny NL}}$, see fig. \ref{fig:w1-wd}(a). 
The polarization enhancement sharply rises with the increase of the drive strength,
reaches a maximum value and then decreases. 
As such, a maximum enhancement in nuclear polarization is observed for a
particular value of drive strength.
We observe a similar trend when we vary
$\omega_{\text{\tiny d}}$ and plot the corresponding enhancement in nuclear polarization
for different $\omega_{\text{\tiny NL}}$, see fig. \ref{fig:w1-wd}(b). Also, the optimal
$\omega_{1}$ is larger for smaller $\omega_{\text{\tiny NL}}$ and vice versa while, the
optimal $\omega_{\text{\tiny d}}$ is smaller for smaller $\omega_{\text{\tiny NL}}$. In
both cases, a smaller $\omega_{\text{\tiny NL}}$ gives higher nuclear polarization
enhancement.


\section{Discussion} 
	
\begin{figure}[t!]
	\centering
	\includegraphics[scale=0.85]{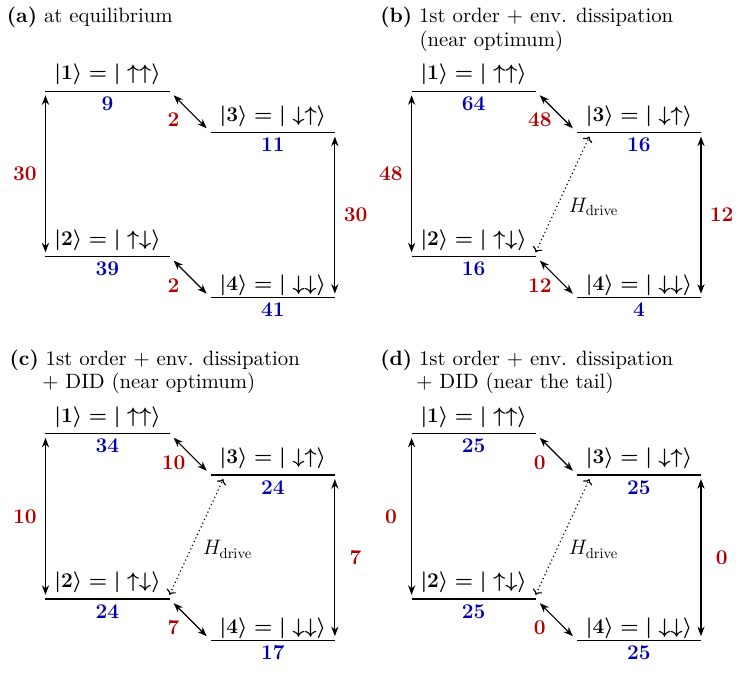}
	\caption{The figure shows the simplified population of different energy levels in
		a coupled electron-nuclear system for various cases. (a) at equilibrium, (b) when the
		first order contribution and environmental dissipation are included, (c) when
		drive-induced dissipation is added in addition to the first order contribution and
		environmental dissipation and optimal value of drive strength is used from fig.
		\ref{fig:w1-wd}a, (d) when drive-induced dissipation is added in addition to the first
		order contribution and environmental dissipation and sub-optimal value of drive strength
		is used from the tail of fig. \ref{fig:w1-wd}a. Here, the drive ($H_{\text{drive}}$) is
		applied at the zero-quantum transition (\textit{i.e.} levels 2-3), and the population is
		approximately hundred times the diagonal elements of the steady-state density matrix. The
		parameters used are $\omega_{n}/2\pi = 300$\,MHz, $\omega_{e} =  10^{3} \times
		\omega_{n}$, $\omega_{\text{d}}/2\pi =  3$\,MHz, $\theta = \pi/3$, $\phi = 0$,
		$\omega_{1}/2\pi =  8$\,MHz (for b, c), $\omega_{1}/2\pi =  80$\,MHz (for d),
		$\omega_{\text{\tiny EL}}/2\pi= 10$\,MHz, and $\tau_{c} = 1$\,ns.}
	\label{fig:Population}
\end{figure}

To physically understand the reason for optimal behavior, we plot the diagonal elements of
the steady-state density matrix (population) for different energy levels of the
electron-nuclear coupled system when the drive is applied at zero-quantum transition
frequency by scaling it a hundred times in fig. \ref{fig:Population}. This is a simplified
picture as it ignores the off-diagonal terms of the steady-state density matrix
(coherence). This helps in intuitively understanding the underlying physics without the
loss of generality. For the notation, $\vert \alpha \beta \rangle$ with $\{\alpha, \beta\}
\in \{\uparrow, \downarrow\}$, given in the figure we have considered that $\alpha$ and
$\beta$ belongs to nuclear and electron spin, respectively. The difference in population
between levels 1-2 and levels 3-4 gives electron polarization, whereas the difference in
population between levels 1-3 and levels 2-4 gives nuclear polarization. 

The fig. \ref{fig:Population}(a) shows the population of different energy levels at
equilibrium. The electron's polarization resulting from both contributions (i.e. levels
1-2 and levels 3-4) is equal. The same is true for nuclear polarization arising from both
the contributions (i.e. levels 1-3 and levels 2-4). When we consider the first order
contribution of the drive on the electron and the dipolar coupling along with the
environmental dissipation (see fig. \ref{fig:Population}(b)), the electron as well as
nuclear polarizations from different contributions becomes unequal. Here, primarily two
effects are competing against each other. The drive is applied to the electron at the
zero-quantum transition between energy levels 2 and 3, to saturate the population of these
levels. On the other hand, the environmental relaxation terms strive to establish a
Gibbsian population ratio between the energy levels corresponding to their thermal
equilibrium values. This intuitively explains the origin of the optimal behavior of
nuclear polarization. 

Furthermore, we incorporate drive-induced dissipation into the system, which opens up new
relaxation pathways. These new pathways allow more population leakage between different
energy levels, as shown in fig. \ref{fig:Population}(c). Thereby further decreasing the
nuclear polarization enhancement. Here, we have considered the optimal drive strength as
given in fig. \ref{fig:w1-wd}(a). When we consider the drive strength from the tails of
the fig. \ref{fig:w1-wd}(a), we notice that the electron as well as nuclear polarization
vanishes, see fig. \ref{fig:Population}(d). This effect becomes prominent when using
stronger drives on the system. This remains the principal difference between our approach
and the existing standard approaches based on Bloch equations.

\begin{figure}[t!]
	\centering
	\includegraphics[scale=0.85]{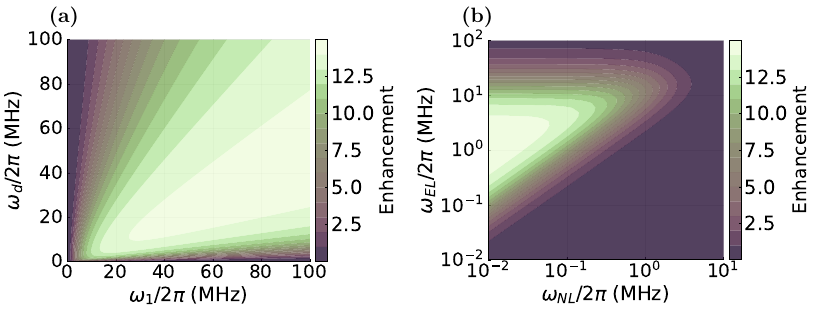}
	\caption{The region of optimality in the parameter space of (a) drive strength
		($\omega_{1}$) and dipolar coupling strength ($\omega_{\text{\tiny d}}$) (b) nuclear and
		electronic subsystem's coupling strength ($\omega_{\text{\text{\tiny NL}}},\,
		\omega_{\text{\text{\tiny EL}}}$) with local environment is shown. The parameters used are
		$\omega_{n}/2\pi = 300$\,MHz, $\omega_{e} =  10^{3} \times \omega_{n}$,
		$\omega_{\text{d}}/2\pi =  3$\,MHz (for b), $\theta = \pi/3$, $\phi = 0$, $\omega_{1}/2\pi
		=  8$\,MHz (for b), $\omega_{\text{\tiny EL}}/2\pi= 10$\,MHz (for a), $\omega_{\text{\tiny
				NL}}/2\pi = 0.1$\,MHz (for a), and $\tau_{c} = 1$\,ns.}
	\label{fig:contour}
\end{figure}

To better understand the correlation between the optimal values of $\omega_{1}$ and
$\omega_{\text{\tiny d}}$, we use a contour plot as shown in fig. \ref{fig:contour}(a).
The contour shows that a positive correlation exists between $\omega_{1}$ and
$\omega_{\text{\tiny d}}$, increasing $\omega_{1}$ results in higher $\omega_{\text{\tiny
d}}$ and vice versa. It is evident that the region of lighter shade remains the optimal
region for solid effect for a given $\omega_{\text{\text{\tiny NL}}}$. 

We used only the part of the dipolar coupling that contributes to the first order. Other
terms do contribute to the second order but not in the first order. In any case, the
optimal behavior due to the drive or the dipolar interaction is not affected by the
presence or the absence of these additional terms.


We take a clue from fig. \ref{fig:w1-wd} which shows that the smaller the value of
$\omega_{\text{\tiny NL}}$, the better the enhancement, and illustrate this behavior
quantitatively in fig. \ref{fig:contour}(b). We predict that the best polarization
transfer requires the smallest possible $\omega_{\text{\tiny NL}}$. The reason is that a
lower $\omega_{\text{\tiny NL}}$ implies a longer relaxation time of the nucleus. This
ensures that the nuclear enhancement is preserved for an extended period.  We also notice
that there exists an optimal electron's coupling strength with the environment. Larger
$\omega_{\text{\tiny EL}}$ implies shorter relaxation time for the electron. In order to
transfer polarization and to preserve it, the relaxation time of the electron must be
orders of magnitude shorter than that of the relaxation time of the nucleus. However, if
$\omega_{\text{\tiny EL}}$ is much higher, the electron will relax before the polarization
is transferred. As such, we predict $\omega_{\text{\tiny EL}}$ also has an optimal value.
The contour plot as shown in fig. \ref{fig:contour}(b) gives us a region of optimality for
$\omega_{\text{\tiny NL}}$ and $\omega_{\text{\tiny EL}}$. If we choose the value of
coupling strength from the region of lighter shade, we can get the maximum possible
enhancement. 

We note that the analysis above was done by applying the drive at the zero quantum
transition. Applying the drive at a double quantum transition or at a single
quantum transition results in the similar optimal behavior.
As such, the theory is consistent for both the two-spin mechanisms of DNP,
\textit{i.e.} the solid effect and Overhauser effect. We note that this model can be easily extended to
cross effect \cite{lily2017} and, with some modifications, to thermal mixing
\cite{lily2017}. 

\section{Conclusion} We have presented a unified description of both the two-spin
mechanisms of dynamic nuclear polarization within a single
theoretical framework using the fluctuation-regularized quantum master equation. Our
analysis reveals the optimal behavior of DNP with respect to the microwave drive strength
and several other parameters for both the two-spin mechanisms of DNP, owing to the
addition of drive-induced dissipation in the dynamics and other higher-order processes. We
demonstrate that exceeding a limit can be detrimental to performance. Hence, an optimal
behavior emerges. We envision these predictions as a means to design better
hyperpolarization protocols. 

\bibliographystyle{apsrev4-1}
\bibliography{DNP_EPL.bib}

\end{document}